\documentclass[a4paper,11pt]{article}
\usepackage{pos}

\title{Multi-wavelength temporal and spectral study of PKS 0402-362}
 \ShortTitle{PKS 0402-362}

\author*[a]{Avik Kumar Das}
\author[b]{Sandeep Kumar Mondal}
\author[c]{Raj Prince}

\affiliation[a]{Indian Institute of Science Education and Research Mohali,\\
Department of Physical Sciences, Knowledge City, Sector 81, SAS Nagar, Punjab 140306, India}

\affiliation[b]{Raman Research Institute, \\
Astrononomy \& Astrophysics department, C. V. Raman Avenue, Sadashivnagar, Bangalore: 560080, India}

\affiliation[c]{Center for Theoretical Physics,  \\
Polish Academy of Sciences, Al.Lotnikow 32/46, 02-668, Warsaw, Poland}

\emailAdd{avikdas@iisermohali.ac.in}
\abstract{We study the long-term behavior of the bright gamma-ray blazar PKS 0402-362. Over a span of approximately 12.5 years, from August 2008 to January 2021, we gathered Fermi-LAT temporal data and identified three distinct periods of intense $\gamma$-ray activity. Notably, the second period exhibited the highest brightness ever observed in this particular source. We observed most of the $\gamma$-ray flare peaks to be asymmetric in profile suggesting a slow cooling time of particles or the varying Doppler factor as the main cause of these flares. The $\gamma$-ray spectrum is fitted with power-law and log-parabola models, and in both cases, the spectral index is very steep. The lack of time lags between optical-IR and $\gamma$-ray emissions indicates the presence of a single-zone emission model. Using this information, we modeled the broadband SEDs with a simple one-zone leptonic model using the publicly available code `GAMERA'. The particle distribution index is found to be the same as expected in diffusive shock acceleration suggesting it as the main mechanism of particle acceleration to very high energy up to 4 - 6 GeV. During the different flux phases, we observed that the thermal disk dominates the optical emission, indicating that this source presents a valuable opportunity to investigate the connection between the disk and the jet.}
\ConferenceLogo{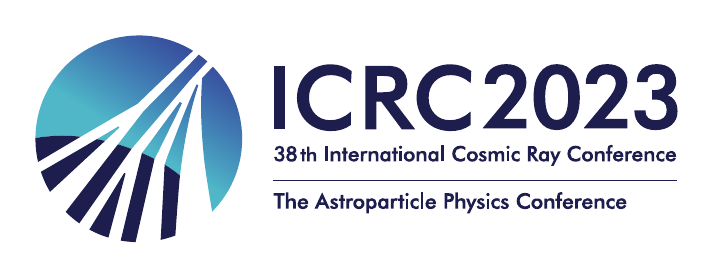}

\FullConference{%
38th International Cosmic Ray Conference (ICRC2023)\\
  26 July - 3 August, 2023 \\
  Nagoya, Japan}

\begin{document}
\maketitle

\section{Introduction}
Active Galactic Nuclei (AGN) are widely accepted to have a supermassive black hole (SMBH) at the center, which accretes matter from the surrounding and forms an accretion disk. A small fraction of these (10 - 15\%) also releases matter in the form of jets perpendicular to the disk plane. AGN is classified as radio galaxies and Seyfert galaxies, depending upon the orientation of the jet to the observer \citep{Urry1995}. In some cases, AGN jets directly point to Earth observers, which are known as blazars, the most powerful subclass of AGNs. The entire emission of blazars is primarily from the relativistic jet and exhibits a characteristic broad double-humped spectral energy distribution (SED) with one peak occurs in the near-infrared (NIR) to X-ray range (low-energy hump), while the other appears at MeV-GeV energies (high-energy hump). The origin of the low-energy hump is widely acknowledged as synchrotron emission from relativistic electrons within the jet. However, the origin of the high-energy hump remains a subject of debate, with arguments pointing towards both leptonic and/or hadronic processes. In the leptonic scenario, high-energy leptons scatter low-energy photons, which may be the synchrotron photons themselves (known as Synchrotron Self-Compton\citep{Maraschi1992}) or can be supplied from outside the jet such as from the Broad-Line Region (BLR), Dusty Torus (DT) or accretion disk (referred to as external Compton \citep{Sikora1994}).  \\
Over the past 15 years, the field of gamma-ray astronomy has experienced significant growth, largely attributed to the remarkable contributions of the Fermi-LAT instrument. The spectral and temporal data from this instument, along with data from other instruments in the optical to x-ray range, presents a remarkable opportunity for researchers to understand the emission region of $\gamma$-ray bright blazars. These include the primary mechanisms responsible for gamma-ray production, the processes behind particle acceleration to exceptionally high energies, the specific locations within the jet where emissions originate, the correlations between different emissions, etc. In this work, our study focuses on addressing some of these inquiries through a comprehensive analysis of the temporal and spectral characteristics of the blazar PKS 0402-362 (4FGL J0403.9-3605). This source is an FSRQ type source located at redshift, $z$ = 1.4228 \citep{Jones2009} with RA = 60.975, Dec = -36.087 and it has been continuously monitored by Fermi-LAT since the beginning of the operation.  %On the other hand, the hadronic model involves proton synchrotron \citep{Mücke2001} or proton-photon \citep{Mannheim1992} or proton-proton interactions \citep{Bednarek1997}.  Depending on the strength of the optical emission lines, blazars are classified into two broad classes namely flat spectrum radio quasars (FSRQ) and BL Lacertae (BL Lacs) objects based on the presence or absence of strong emission lines in their optical spectra, respectively. The most recent publication of the Fermi catalog (4FGL DR3) has identified a large number of gamma-ray sources, with approximately 90\% of them classified as blazars.

%In this paper, we have analyzed the Fermi-LAT data collected over a period of 12 years from this source (ref section \S\ref{sec:2}) and identified the flares in its long-term light curve (ref section \S\ref{sec:3}). The gamma-ray SED and correlation for all the episodes are presented in section \S\ref{sec:4} and section \S\ref{sec:5}, respectively. Subsequently, we discussed the broadband SED modeling in section \S\ref{sec:6} and detailed discussions in section \S\ref{sec:7}.

%The central part (SMBH+disk+base of jets) of the AGN is inferred to be surrounded by an optically thick molecular gas cloud called the molecular torus or dusty torus.

\section{DATA ANALYSIS} \label{sec:2}

\subsection{Gamma-ray}
%Fermi-LAT (Large Area Telescope) is an imaging pair conversion $\gamma$-ray telescope, which covers the energy range from 50 MeV to 1 TeV with a wide field of view of $>$ 2.4 sr \citep{Atwood2009}. The complete characteristics of LAT are provided on the Fermi Webpage \footnote{\url{https://fermi.gsfc.nasa.gov/ssc/data/analysis/software/}}. 
%It has an orbital period of 96 minutes and scans the entire sky every two orbits ($\sim$ 3.2 hours).

We analyzed the Fermi-LAT\footnote{\url{https://fermi.gsfc.nasa.gov/ssc/data/analysis/software/}} data for the gamma-ray observation. We downloaded the Pass8R2 data from August 2008 to January 2021 (approximately 12.5 years) from the blazar, PKS 0402-362. This data was analyzed using the Fermi science tool (version 1.0.10). We used `evclass=128' and `evtype=3' with energies ranging from 100 MeV to 300 GeV. To avoid gamma-ray contamination from the Earth's limb, we applied a maximum zenith angle cut of 90 degrees. We used the galactic diffuse emission model (\texttt{gll\_iem\_v07.fits}) and isotropic diffuse emission model (\texttt{iso\_P8R3\_SOURCE\_V3\_v1.tx}) as background models, along with other point sources. We employed the preferred method, "Unbinned likelihood analysis", and kept all the source parameters fixed that were away from the Region of Interest (ROI) of 10 degrees to extract the light curve and SED.
%We have further followed the same procedure for the analysis as given in \cite{2020ApJS..248....8D}. To select the good time interval data, we implemented the filter expression ``\texttt{DATA\_QUAL}$>$0 \&\& \texttt{LAT\_CONFIG}==1 \&\& \texttt{ANGSEP}(\texttt{RA\_SUN},\texttt{DEC\_SUN},60.974,-36.0839)$>$15".

%\begin{figure*}
%\centering

%\includegraphics[height=2.6in,width=5.3in]{figures/flare1_bb.pdf}
%\label{fig:2}

%\includegraphics[height=2.6in,width=5.3in]{figures/flare1_fitted.pdf}
%\caption{\textbf{Upper panel}: Two-day binning light curve of AE-1A. The time duration of the different phases is MJD 55176-55216 (Pre-flare), MJD 55216-55256 (flare-1A1), MJD 55256 - 55276 (flare-1A2), and MJD 55276-55334 (Post-flare). These are shown by a dash-dot magenta line. \textbf{Lower panel}: The fitted light curve of flare-1A1 and flare-1A2 with a time span of 40 days (MJD 55216 - 55256) and 20 days (MJD 55256 - 55276), respectively. Here, the light curve of both phases is shown in one figure.}
%\label{fig:2}
%\end{figure*}
 
\subsection{X-ray and optical}
We analyzed the X-ray and ultraviolet-optical data from the Neil Gehrels Swift Observatory (Swift-XRT/UVOT), which were taken during the same period as the gamma-ray observations from the HEASARC webpage\footnote{\url{https://heasarc.gsfc.nasa.gov/cgi-bin/W3Browse/swift.pl}}. We chose a circular radius of 25 arc seconds for the source region and an annular ring around the source as the background region. The grouped spectra were fitted in XSPEC (version 12.11.0) using two models: $powerlaw$' and $logparabola$'. The log parabola model was selected as the best-fitting model if the F-test probability was less than or equal to 0.05. To account for galactic absorption, we used the `$tbabs$' model with a neutral hydrogen column density ($n_{H}$) of $6.92\times10^{19} cm^{-2}$ as provided on the HEASARC webpage\citep{HI4PI2016}. 
%Due to low photon counts, we grouped the spectra with at least one count per bin for a few observation ids. We employed C-statistics and used the `$powerlaw$' model to fit those cases' spectra. `$xrtmkarf$' and `$grppha$' has been used to create the ancillary response file and group the spectra, respectively. Next, The task `$xrtpipeline$' (version 0.13.2) was used to process the XRT data \citep{Burrows2005} with an energy range from 0.3 to 10.0 keV for each observation ID. This process implemented the calibration files with version 20160609 and standard screening criteria.

For the UV-optical observation, we analyzed the data from UVOT \citep{Roming2005}. For each of the six filters (V, B, U, W1, M2, and W2), a circular region with a radius of 5 arc seconds was chosen to analyze the data, while an annular region surrounding the source was selected as the background region. The task known as 'uvotsource' was utilized to extract the magnitudes of the source. After accounting for galactic reddening \citep{Schlafly2011} and atmospheric extinction, the magnitudes were converted into flux values using the appropriate zero points \citep{Breeveld2011} and conversion factors \citep{Larionov2016}.

We also utilized publicly available archival data from SMARTS and the Steward Optical Observatory, both part of the {\it Fermi} Blazars monitoring program. SMARTS provide BVRJ photometric data\citep{Bonning2012}, while Steward Observatory offered V and R band photometric data, along with polarimetric data\citep{Smith2009}.
%\subsection{SMARTS and Steward Data}

%We utilized publicly available archival data from SMARTS\footnote{\url{http://www.astro.yale.edu/smarts/glast/home.php}} and the Steward Optical Observatory\footnote{\url{http://james.as.arizona.edu/psmith/Fermi/}}, both part of the {\it Fermi} Blazars monitoring program. SMARTS provide BVRJ photometric data\citep{Bonning2012}, while Steward Observatory offered V and R band photometric data, along with polarimetric data\citep{Smith2009}. 
%, followed by the usage of $uvotimsum$' to combine multiple observations
%We have collected BVRJ data of this source during MJD 55838 - 57297.

\section{Gamma-ray Light curves} \label{sec:3}
Based on the simultaneous observations from Swift-XRT/UVOT, we identified three prominent Activity Epochs (AE) in the $\gamma$-ray light curve: AE-1, AE-2, and AE-3. These epochs were analyzed in one day binning for further study. We employed Bayesian Block (BB) \citep{Scargle2013} representation \footnote{\url{https://docs.astropy.org/en/stable/api/astropy.stats.bayesian\_blocks.html}} in this finer binned (one-day or two-day) light curve to identify different sub-structures: AE-2A, AE-2B, AE-3A, etc. Finally, these sub-structures were classified into different phases, such as Pre-flare, flare, Post-flare, etc., based on the flux levels derived from the BB algorithm. To determine the duration of the flare phase, we utilized Ivan Kramarenko's algorithm \citep{Geng2020}, which employs an iterative approach to segregate data points into two sets in order to calculate a threshold value. The `flare' phase is considered when the flux level exceed the threshold value.

% We conducted an analysis of the Fermi-LAT data for PKS 0402-362 spanning a period of 12.5 years (MJD 54682 - 59220) and binned it into intervals of 7 days. The threshold value is defined as $<F> + 2 \times F_{Disp}$, where $<F>$ is the mean flux value of the light curve and $F_{Disp}$ is the true dispersion value of the low flux state. More details of the algorithm can be found in \citet{Geng2020Nov}.
Each flare phase's light curve consists of several peaks. We fitted these peaks with the sum of exponential function to compute the rising ($T_r$) and decay time ($T_d$) of each peak: $F(t)=2\sum_{i}^{n}F_{0,i}\Big[\exp \Big(\frac{t_{0,i}-t}{T_{r,i}}\Big)+\exp \Big(\frac{t-t_{0,i}}{T_{d,i}}\Big)\Big]^{-1}$, Where, $i$ runs over the number of major peaks (n).$F_0$ is the photon flux at time $t_{0}$ for a particular peak. We used Bayesian Information Criteria or BIC value to choose the reasonable number of exponential functions to fit the light curve. There are four parameters per exponential function in our case. We chose one, two, or three exponential functions together to fit the light curve of the flare phase for which the $BIC$ value is minimum. In this draft, all the reported $\gamma$-ray photon fluxes are in the unit of $10^{-6}$ ph cm$^{2}$ s$^{-1}$. 

\section{Study of $\gamma$-ray SEDs} \label{sec:4}
We analyzed the gamma-ray spectral energy distributions (SEDs) for different phases (such as Pre-flare, flare, Post-flare, etc). Two different models: the power-law and log-parabola models, were used to fit these SEDs. In order to compare these two models, we have computed the log-likelihood values for different phases. Based on the differences in log-likelihood ($\Delta log\mathcal{L} = \lvert log\mathcal{L}{LP} \rvert - \lvert log\mathcal{L}{PL} \rvert$), we found that in most cases, the gamma-ray SEDs were better described by the log-parabola model over the power-law model. We also found that for most cases (16 out of 22 cases), the gamma-ray SEDs exhibited a `spectral hardening' behavior. This means that the spectral index decreased (increased) when the source transitioned from a low to high (high to low) flux state. We observed only one case where the source showed a significant `spectral softening' behavior. In other five cases, the spectral index remained nearly constant within the uncertainties.
%1. A power-law model: $\frac{dN}{dE} = N_{0}\Big(\frac{E}{E_0}\Big)^{-\Gamma_{PL}}$

%\begin{equation}
%   \frac{dN}{dE} = N_{0}\Big(\frac{E}{E_0}\Big)^{-\Gamma_{PL}}
%\end{equation}

%where, $N_{0}$, and $\Gamma_{PL}$ are the normalization factor and spectral index of the model. We have kept free these parameters during likelihood fitting, while the scaling factor ($E_{0}$) is fixed at 100 MeV for all the cases.

%2. A log-parabola model: $\frac{dN}{dE} = N_{s}\Big(\frac{E}{E_s}\Big)^{-(\alpha+\beta\log(E/E_s))}$,

%\begin{equation}
%   \frac{dN}{dE} = N_{s}\Big(\frac{E}{E_s}\Big)^{-(\alpha+\beta\log(E/E_s))}
%\end{equation}

%This model is similar to the power law but with an energy-dependent photon index ($a(E) = \alpha + \beta \log(E/Es)$). $\alpha$ is the photon index at energy $E_s$, which is fixed at 300 MeV, near the low energy part of the spectrum. $\beta$ is known as the curvature index of the model.

%In order to compare these two models, we have computed the log-likelihood values for different phases. From The differences in log-likelihood ($\Delta log\mathcal{L} = \lvert log\mathcal{L}_{LP} \rvert - \lvert log\mathcal{L}_{PL} \rvert$), we found for most of the cases, the $\gamma$-SEDs are best described by the Log-parabola over Power-law model. %The values of different model parameters and the negative log-likelihood are given in Table-\ref{tab:1} for AE-1A.

\section{CORRELATION STUDY} \label{sec:5}

A correlation study was conducted between $\gamma$-ray and various optical bands (SMARTS- B, V, R, and J bands) for AE-2 and AE-3, respectively. However, due to the limited data points in the X-ray band, we can not examine the correlation between $\gamma$-ray and X-ray data. 

Unbinned Discrete Correlation Function (UDCF) is given by \citep{Edelson1988} $UDCF_{ij} = \frac{(a_i - <a>)(b_j - <b>)}{\sqrt{ (\sigma_a^2 - e_a^2)(\sigma_b^2 - e_b^2)}}$. where, $a_i$, and $b_j$ are the two discrete time series. $\sigma_a$ and $\sigma_b$ are the standard deviations of the two-time series, respectively. Here, each value of $UDCF_{ij}$ is associated with the time delay, $\Delta t_{ij} = t_{j} -  t_{i}$. If we average this equation over M number of pairs for which $\tau - \Delta \tau/2 \leq \Delta t_{ij}  (= t_j - t_i) \leq \tau + \Delta \tau/2$, we get, the discrete correlation function ($DCF(\tau)$):

\begin{equation}
    DCF(\tau) = \frac{1}{M} \sum UDCF_{ij}  \pm  \sigma_{DCF}(\tau)  
\end{equation}

$\sigma_{DCF}(\tau)=\frac{1}{M-1} \sqrt {\sum [UDCF_{ij} - DCF(\tau)]^2}$ is the associated error at the time lag $\tau$.

The correlation between $\gamma$-ray and all the four optical bands (B, V, R, J) for AE-2 show minimal negative time lag of -2.50, -4.00, -4.00 and -2.50 days with DCF values of 0.70$\pm$0.19, 0.71$\pm$0.23, 0.73$\pm$0.24, and 0.71$\pm$0.31 respectively. For epoch AE-3, the DCF plots between $\gamma$-ray and optical bands show very small positive time lag (1.50, 2.00, 2.50, and 2.00 days) with high DCF values of 0.79$\pm$0.21, 0.73$\pm$0.16, 0.71$\pm$0.15, and 0.75$\pm$0.15, respectively. This suggests that light curves of $\gamma$-ray and optical bands are positively correlated. Here, Negative lag implies that the $\gamma$-ray light curve lags the optical emission. We also checked the correlation study results with ICCF method\footnote{\url{http://ascl.net/code/v/1868}} and computed uncertainty on time lags by Monte Carlo (MC) analysis which involves Flux Randomization (FR) and Random Subset Selection (RSS) techniques \citep{Peterson1998}. Time lags obtained by ICCF are consistent with DCF results within the error bar.

%To estimate the significance level of the DCF peaks, we have followed the procedure given in \citep{Max-Moerbeck2014}. In Figure-\ref{fig:5}, 90\%, 95\%, and 99\% (from light green to deep green) significance levels are shown for AE-2 (B-gamma) and AE-3 (B-gamma) cases. We found that all the combinations show a strong correlation above 99$\%$.
%After modeling the observed periodograms using the `PSRESP' method \citep{Uttley2002May}, we have simulated 1000 light curve pairs by \citep{Timmer1995} algorithm for each case. Finally, we cross-correlate the simulated light curve pair to estimate the significance level for each time lag. The results of the correlation study are shown in Figure-\ref{fig:5}. All the figures on the left and right columns are for AE-2 and AE-3 epochs, respectively.

%However, for both epochs, it is important to note that the time lags seen here in all the combinations are shorter than the average time gap of the optical light curve ($\sim$5 days for AE-2 and $\sim$3 days for AE-3). The observed small time lags could be caused by the gap in the light curves and, in that case, can not be considered reliable. For the AE-2 epoch, we found centroid time lags of -3.1$^{+1.9}_{-8.9}$, -2.9$^{+2.9}_{-6.8}$, -3.0$^{+1.9}_{-2.0}$, and -11.9$^{+13.9}_{-1.0}$ days for B-gamma, V-gamma, R-gamma, and J-gamma correlation respectively. Whereas, for AE-3 epoch, the value of centroid lags are: 0.05$^{+0.99}_{-0.95}$,  0.04$^{+0.18}_{-0.98}$,  0.04$^{+0.11}_{-0.96}$, and  0.06$^{+0.12}_{-1.67}$ days.
\section{MULTI-WAVELENGTH MODELLING} \label{sec:6}
%\textcolor{red}{In Progress}
We utilized `GAMERA', an open-source code, to model the multi-wavelength SEDs of the blazar PKS 0402-362. This code is publicly available on GitHub\footnote{\url{https://github.com/libgamera/GAMERA}}. This code solves the time-dependent transport
equation; it estimates the propagated particle spectrum N(E,t) for an input injected particle spectrum. It further uses the propagated spectrum to calculate the Synchrotron and Inverse-Compton (IC) emissions. GAMERA solves the following transport equation:
 \begin{equation}
\frac{\partial N(E,t)}{\partial t}= Q(E,t)-\frac{\partial}{\partial E}(b(E,t)N(E,t))-\frac{N(E,t)}{\tau\textsubscript{esc}(E,t)}
\end{equation}
where, Q(E,t) is the input particle spectrum and b(E,t) corresponds to the energy loss rate by Synchrotron and IC and can be defined as
\big($\frac{dE}{dt}$\big). \\
We used LP model as a particle distribution to model the multi-wavelength SEDs because the LP model gives the best-fit parameters for $\gamma$-ray spectrum in most of the phases. Following \cite{Massaro2004}, an LP photon spectrum can be produced by the radiative losses of an LP electron spectrum.
%\begin{equation}
%Q(E)=L\textsubscript{o}\Bigg(\frac{E}{E_o}\Bigg)^{-\big(\alpha+\beta log\big(\frac{E}{E_o}\big)\big)}
%\end{equation}
%where L\textsubscript{o} is the normalization constant and E\textsubscript{o} is the scaling factor. 
%This code uses the `Klein-Nishina' cross-section to compute Inverse Compton emission \citep{Blumenthal1970Jan}.

\begin{figure*}
\centering

\includegraphics[height=2.2in,width=2.9in]{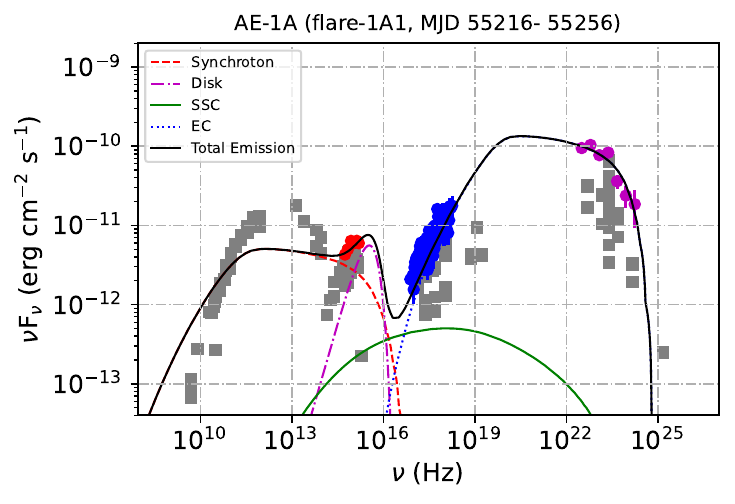}
\includegraphics[height=2.2in,width=2.9in]{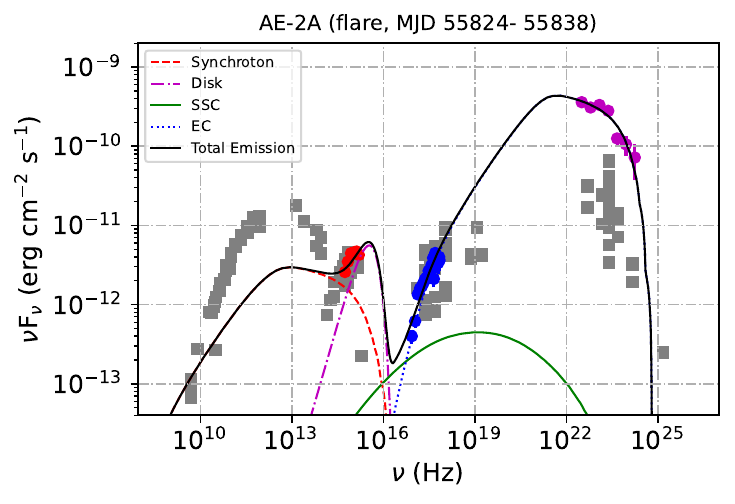}
\caption{Simple one-zone model fits the multi-wavelength SEDs of two phases (flare-1A1 of AE-1A and flare of AE-2A). Swift-UVOT and XRT data are shown in red and blue colors respectively. Magenta color points represent the $\gamma$-ray data.}

\label{fig:1}
\end{figure*}

\subsection{Physical Constraints}
PKS 0402-362 is an FSRQ-type object. During the IC process, besides SSC (Synchrotron self-Compton) we consider External Compton (EC) process in which the seed photons are coming from Broad Line Region (BLR) and accretion disk. Here we have discussed the physical constraints on the model parameters that we have used in our modeling:

1. To calculate the EC emission by the relativistic electrons, Broad Line Region (BLR) photons are taken into account as target photons. The energy density of the BLR photons has been calculated (in the co-moving frame) with the equation: $ U_{BLR}^{\prime}=\frac{\Gamma^2 \zeta_{BLR}L_{Disk}}{4\pi cR^2_{BLR}}$, where $\Gamma$ is the bulk Lorentz factor of the emitting blob whose value is assumed to be 16.24 \citep{Hovatta2009}. The BLR photon energy density is only a fraction of 10\% ($\zeta_{BLR}\sim$ 0.1) of the accretion disk photon energy density. The assumed value of the disk luminosity $L_{Disk}$= 5$\times$10$^{46}$ erg/sec, which is a typical value of FSRQ-type blazars. The radius of the BLR region is derived from the scaling relation, $R_{BLR} \sim (\frac{L_{Disk}}{10^{45}})^{\frac{1}{2}} \times 10^{17}$ cm \citep{Ghisellini2009}.
        
2. We also include the effect of EC emission for accretion disk photons. We computed the disk energy density in the co-moving jet frame by the equation \citep{Dermer2009}: $U_{Disk}^{\prime}=\frac{0.207R_g L_{Disk}}{\pi c Z^3 \Gamma^2}$. The mass of the central engine or Black Hole (M$_{BH}$) is 10$^9$M$_{\odot}$. So, the gravitational radius R$_g$=3$\times$10$^{14}$ cm (gravitational radius of sun= 3$\times$10$^{5}$ cm.). The distance of the emission region from the black hole is given by Z$\le \frac{2 \Gamma^2 c t_{var}}{1+z}$. Where, the variability time ($t_{var}$) is estimated as $\sim$ 1.18 days by the procedure described in \citep{Das2020}. 
%This gives the value of `Z':6.7$\times$10$^{17}$ cm.
%        \item In the modeling, we consider a spherical emission region or blob of radius R. The size of the emission region (R) can be constrained from the following equation:
%        \begin{equation} \label{eq:14}
%            R\le\frac{c t_{var} \delta}{1+z}
%        \end{equation}
%        where t$_{var}$ is the observed variability time, $\delta$ is the Doppler factor of the emission region, which is assumed as 17.0 \citep{Hovatta2009Feb}. z is the redshift of the source. This gives just an approximate constraint on the size of the emission region, as there are several other factors that may affect this estimate  \citep{2002PASA...19..486P}.

3. The Optical-UV part of the broadband SED is explained by the multi-temperature accretion disk model with the temperature profile given as: $T(R) =  \Big[\frac{3 R_{s} L_{Disk}}{16 \pi \eta \sigma R^{3}} \Big(1 - \sqrt{\frac{3 R_{s}}{R}} \Big) \Big]^{\frac{1}{4}}$. Here, $R_{s}$ (= $\frac{2 G M}{c^{2}}$) is Schwarzchild radius. The maximum flux of the accretion disk spectrum occurs at $\sim 5 R_{s}$ \citep{Ghisellini2009mnras}. $\sigma$ and $\eta$ are the Stefan-Boltzman constant and efficiency ($L_{Disk} = \eta \dot{M} c^{2}$) of the accretion disk, respectively.\\
After constraining the above model parameters, we model the multi-wavelength SEDs (five activity phases) of PKS 0402-362 by varying following free parameters: magnetic field (B), the radius of the emission region (R), minimum and maximum electron energy (e$_{min}$ and e$_{max}$), spectral index ($\alpha$) and curvature index ($\beta$). We assumed a constant escape of leptons from the emission region with an escape time of $\tau_{esc} \sim$ R/c, where c is the speed of light. All the values of the fitted parameters for various phases are given in Table-\ref{tab:1}. \\
\begin{table*}
\caption{Results of multi-wavelength SED modeling (One-zone). Second to fifth columns represent the value of various parameters used in modeling. Here,  $e_{min}$, $e_{max}$ = Minimum and maximum energy of injected electrons, R = Size of the emission region ($\le\frac{c t_{var} \delta}{1+z}$), $P_e$ = Power in the injected electrons, $P_B$ = Power in the magnetic field, $P_{tot}$ = Total required jet power. Several parameters are kept fixed at a specific value during modelling/fitting: $\alpha$, $\beta$ = Spectral and Curvature index of injected electron spectrum = 2.00, 0.09, $T^\prime_{BLR}$ = 2.0$\times10^{4}$ K, $T^\prime_{Disk}$ = 1.0$\times10^{6}$ K, $U^\prime_{BLR}$ = 7.14 erg/$cm^3$, $U^\prime_{Disk}$ = 2.08$\times10^{-7}$ erg/$cm^3$, $\delta$ = 17.0, $\Gamma$ = 16.24.}
\label{tab:1}
\centering
\begin{tabular}{ccccccc rrrr}   % creating eight columns
\hline\hline
Activity & $e_{min}$ & $e_{max}$ & B & R & $P_e$ & $P_B$ & $P_{tot}$ \\ %& Time duration  \\
& (MeV) & (MeV) & (G) &(cm.) & (erg/sec) & (erg/sec) & (erg/sec) \\ %& (days) \\
\hline\hline
AE-1A & 24.53  & 5110.00 & 2.6 & 4.0$\times10^{16}$ & 2.37$\times10^{46}$ & 1.07$\times10^{46}$  & 4.35$\times10^{46}$ \\ %& 40 \\
(flare-1A1) & \\
\hline
AE-1A & 24.53 & 6132.00 & 2.4 & 4.0$\times10^{16}$ & 2.45$\times10^{46}$  & 9.11$\times10^{45}$ & 4.23$\times10^{46}$ \\ %& 20 \\
(flare-1A2) & \\
\hline
AE-2A & 132.86 & 4854.50 & 1.1 & 5.0$\times10^{16}$ & 4.34$\times10^{46}$  & 2.99$\times10^{46}$ & 5.09$\times10^{46}$ \\ %& 14  \\
(flare) & \\
\hline
AE-3A & 76.65 & 4854.00 & 2.8 & 6.5$\times10^{16}$ & 1.36$\times10^{46}$  & 3.28$\times10^{45}$  & 4.84$\times10^{46}$ \\ %& 17 \\
(flare) & \\
\hline
AE-3D & 48.54 & 3832.00 & 2.8 & 4.0$\times10^{16}$ & 2.18$\times10^{46}$  & 1.24$\times10^{46}$   & 3.93$\times10^{46}$ \\  %& 60  \\
(Post-flare) & \\
\hline\hline
\end{tabular}
\label{tab:MWSED_Param}
\end{table*}
\section{Discussion and Conclusions} \label{sec:7} 
Based on the $\sim$ 12.5 years of $\gamma$-ray light curve the source has been observed three times in a major activity phases identified as AE-1, AE-2, and AE-3. Among these epochs, AE-2 was the brightest, with a flux exceeding 20 times the average gamma-ray flux. These epochs were further divided into different sub-epochs, such as AE-1A, AE-2A, etc. Next, We used the Bayesian Block method to identify different activity phases from these sub-epochs: pre-flare, flare, post-flare. Flare phases were found to be a combination of many peaks which were fitted with a standard sum of exponential to estimate the rise and decay timescale (T$_{r}$ and T$_{d}$). To compare these two timescales, we define a asymmetry parameter K = $\frac{T_d - T_r}{T_d + T_r}$. A peak is symmetric if the value of $\lvert K \rvert \leq$ 0.3. In our study, most of the time we found asymmetric peaks either with fast rise slow decay or slow rise fast decay except three cases where it is more symmetric within the error bar. The variability in the flare can be caused by the interaction of the blob (emission region) with a standing shock \citep{Blandford1979}. However, there could be other possibilities, such as magnetic reconnection \citep{Spada1559}. The symmetric peaks are expected to occur when the particles radiate all their energy within the light-crossing time ($\sim$ R/c), and observation of such peaks can be compatible with the scenario of a blob passing through a standing shock. Considering the scenario where the blob interacts with shocks, once the blob is moved out or the shock is moving we expect to see asymmetric peaks. This can also be seen as a scenario where the radiative cooling time scale is longer than the particle injection time scale and as a result, a long-decay flare is observed. \\ %the interaction of internal shocks with moving blob or\\
The optical-IR light curves were available only for the AE-2 and AE-3 phases. We found a nearly zero time lag between the gamma-ray and optical-IR emissions, suggesting their co-spatial origin and demanding a single emission zone in the broadband SED modeling. PKS 0402-362 is classified as an FSRQ-type source, indicating that the disk emission is expected to be higher compared to BL Lac type sources. We observed a dominant accretion disk for this source, which is well-fitted by a combination of synchrotron and thermal disk emissions (see Figure-\ref{fig:1}). The presence of a strong accretion disk makes this source a suitable candidate for exploring the disk-jet connection. Scaling arguments \citep{Ghisellini2009} suggest that the emission region is located at the boundary of the broad-line region (BLR), potentially providing seed photons for external Compton (EC) processes. Consequently, we found that the high-energy peak of the SED is well described by EC emission from the BLR photons. The broadband SED of two phases (i.e., flare-1A1 of AE-1A and the flare phase of AE-2A) is shown in Figure-\ref{fig:1}, along with the archival data shown in grey color. The magnetic field was found to be between 1.1 and 2.8 Gauss, which is within the expected range for blazar. The size of the emission region determined from the fitting process closely agrees with the observation-based estimate derived from the variability time. The calculated jet power in terms of electron and magnetic field contributions ($P_{B}$ and $P_{e}$) are comparable, but the electron power dominates. The total required jet power is well within the
Eddington luminosity of the source. 


\begin{thebibliography}{99}

\bibitem{Urry1995}
Urry C. M., Padovani P., 1995, Publ. Astron. Soc. Pac.

%\bibitem{Goyal2018}
%Goyal A., 2018, Galaxies, 6

%\bibitem{Blinov2015}
%Blinov D., et al., 2015, Mon. Not. R. Astron. Soc., 45

%\bibitem{Avachat2015}
%Avachat S. S., Perlman E. S., Adams S. C., Cara M., Owen F., Sparks W. B., Georganopoulos M., 2016, Astrophys. J.

\bibitem{Maraschi1992}
Maraschi L., Ghisellini G., Celotti A., 1992, ApJ

\bibitem{Sikora1994}
Sikora M., Begelman M. C., Rees M. J., 1994, ApJ, 421, 153

%\bibitem{Mücke2001}
%Mücke A., Protheroe R. J., 2001, Astroparticle Physics, 121

%\bibitem{Mannheim1992}
%Mannheim K., Biermann P. L., 1992, A\&A, L21

%\bibitem{Bednarek1997}
%Bednarek W., Protheroe R. J., 1997, MNRAS, 287

\bibitem{Jones2009}
Jones D. H., et al., 2009, Mon. Not. R. Astron. Soc., 3

%\bibitem{Ballet2020}
%Ballet J., Burnett T. H., Digel S. W., Lott B., 2020, Fermi
%Large Area Telescope Fourth Source Catalog Data Release 2,
%doi:10.48550/ARXIV.2005.11208, https://arxiv.org/abs/2005.
%11208

%\bibitem{Atwood2009}
%Atwood W. B., et al., 2009, ApJ, 697, 1071

\bibitem{Burrows2005}
Burrows D. N., et al., 2005, Space Sci. Rev., 120, 165

\bibitem{HI4PI2016}
HI4PI Collaboration et al., 2016, Astron. Astrophys., 59

\bibitem{Roming2005}
Roming P. W. A., et al., 2005, Space Sci. Rev., 120, 95

\bibitem{Schlafly2011}
Schlafly E. F., Finkbeiner D. P., 2011, ApJ, 737

\bibitem{Breeveld2011}
Breeveld A. A., Landsman W., Holland S. T., Roming P., Kuin N. P. M., Page M. J., 2011, AIP Conf. Proc., 1358

\bibitem{Larionov2016}
Larionov V. M., et al., 2016, MNRAS, 46

\bibitem{Bonning2012}
Bonning E., et al., 2012, Astrophys. J., 75

\bibitem{Smith2009}
Smith P. S., Montiel E., Rightley S., Turner J., Schmidt G. D., Jannuzi B. T., 2009, arXiv e-prints, p. arXiv:0912.3

\bibitem{Scargle2013}
Scargle J. D., Norris J. P., Jackson B., Chiang J., 2013, ApJ, 7

\bibitem{Geng2020}
Geng X., et al., 2020, Astrophys. J., 904, 67

\bibitem{Meyer2019}
Meyer M., Scargle J. D., Blandford R. D., 2019, Astrophys. J.,877, 39

\bibitem{Edelson1988}
Edelson R. A., Krolik J. H., 1988, ApJ, 333, 646

%\bibitem{Max-Moerbeck2014}
%Max-Moerbeck W., et al., 2014, MNRAS, 445, 428

%\bibitem{Timmer1995}
%Timmer J., Koenig M., 1995, Astron. Astrophys., 300, 707

\bibitem{Peterson1998}
Peterson B. M., Wanders I., Horne K., Collier S., Alexander T., Kaspi S., Maoz D., 1998, Publ. Astron. Soc. Pac., 110, 660 

\bibitem{Massaro2004}
Massaro E., Perri M., Giommi P., Nesci R., 2004, A\&A, 413, 489

\bibitem{Hovatta2009}
Hovatta T., Valtaoja E., Tornikoski M., Lähteenmäki A., 2009, Astron. Astrophys., 494, 527

\bibitem{Ghisellini2009}
Ghisellini G., Tavecchio F., 2009, Mon. Not. R. Astron. Soc., 397, 985

\bibitem{Das2020}
Das A. K., Prince R., Gupta N., 2020, ApJS, 248, 8

\bibitem{Dermer2009}
Dermer C. D., Menon G., 2009, High energy radiation from black holes:
gamma rays, cosmic rays, and neutrinos. Princeton series in astrophysics, Princeton Univ. Press, Princeton, NJ, https://cds.cern.ch/record/1225453

\bibitem{Ghisellini2009mnras}
Ghisellini G., Tavecchio F., Ghirlanda G., 2009, Mon. Not. R. Astron. Soc., 399, 2041

\bibitem{Gilmore2012}
Gilmore R. C., Somerville R. S., Primack J. R., Dominguez A., 2012, Mon. Not. Roy. Astron. Soc., 422, 318

\bibitem{Blandford1979}
Blandford R. D., Königl A., 1979, ApJ, 232

\bibitem{Spada1559}
Spada M., Ghisellini G., Lazzati D., Celotti A., 2001, Mon. Not. R. Astron. Soc., 325, 1559


\end{thebibliography}
\end{document}